\documentclass[showpacs,preprintnumbers,amsmath,amssymb,floatfix]{revtex4}
\usepackage{epsf}
 \newcommand{\insertplot}[5]{\begin{figure}
 \hfill\hbox to 0.05in{\vbox to #5in{\vfill
 \inputplot{#1}{#4}{#5}}\hfill}
 \hfill\vspace{-.1in}
 \caption{#2}\label{#3}
 \end{figure}}
 \newcommand{\inputplot}[3]{
 \special{ps: plotfile #1}
}

\usepackage[german, english]{babel}
\usepackage{ifthen}
\usepackage{epsfig}
\newcounter{fig}   \newcommand{\lbfig}[1]{\refstepcounter{fig}
\label{#1} }

\newcommand{\vphi}{\varphi}

\begin{document}
\title{Sphalerons, Antisphalerons and Vortex Rings}
\author{
{\bf Burkhard Kleihaus, Jutta Kunz and Michael Lei\ss ner}
}
\affiliation{
{Institut f\"ur Physik, Universit\"at Oldenburg, 
D-26111 Oldenburg, Germany}
}
\date{\today}
%\pacs{04.20.Jb, 04.40.Nr}

\begin{abstract}
We present new classical solutions of Weinberg-Salam theory
in the limit of vanishing weak mixing angle.
In these static axially symmetric solutions, 
the Higgs field vanishes either on isolated points
on the symmetry axis, or on rings centered around the symmetry axis.
The solutions represent systems of sphalerons, antisphalerons, 
and vortex rings.
\end{abstract}

\maketitle

\section{Introduction}

The standard model does not absolutely conserve baryon and lepton number
\cite{thooft}.
Fermion number violation can arise because of instanton transitions
between topologically inequivalent vacua \cite{thooft,ring}.
At finite temperature,
fermion number violating processes may arise because of 
thermal fluctuations of the fields,
large enough to overcome the energy barrier between topologically
distinct vacua \cite{manton,km}.
The rate for baryon number violating processes
is then largely determined by a Boltzmann factor,
containing the height of the barrier at a given
temperature. 

In Weinberg-Salam theory, 
this energy barrier between topologically inequivalent vacua
is associated with the Klinkhamer-Manton sphaleron \cite{km,kkb},
an unstable classical configuration.
The energy of the sphaleron determines the barrier
height, and its single unstable mode is crucial for the
baryon number violating processes \cite{review}.
The sphaleron itself carries baryon number $Q_{\rm B}=1/2$ \cite{km}.

However, 
the non-trivial topology of the configuration space of Weinberg-Salam theory
gives rise to further unstable classical solutions.
A superposition of $n$ sphalerons, for instance,
can lead to static axially symmetric solutions,
multisphalerons, whose energy density is torus-like \cite{multi}.
These multisphalerons carry baryon number $Q_{\rm B}=n/2$ 
\cite{bk,multi}.
But axial symmetry is not necessary for such multisphaleron
configurations.
Recently, we constructed multisphaleron solutions, which
possess only discrete symmetries, platonic sphalerons
\cite{Kari}.
In particular, we obtained tetrahedral, cubic and octahedral
sphalerons.

Klinkhamer, on the other hand, constructed a static axially
symmetric solution, which may be thought of as a bound
sphaleron-antisphaleron system, in which a sphaleron and an
antisphaleron are located at an equilibrium distance
on the symmetry axis \cite{klink}.
Such a sphaleron-antisphaleron pair has vanishing baryon number,
$Q_{\rm B}=0$, since the antisphaleron carries $Q_{\rm B}=-1/2$.
The sphaleron-antisphaleron pair therefore does not mediate
baryon number violating processes.

We here show that
these sphaleron-antisphaleron pair solutions can be generalized,
to form sphaleron-antisphaleron chains,
where $m$ sphalerons and antisphalerons are located
on the symmetry axis, in static equilibrium,
as conjectured previously \cite{bk}.
These solutions are thus analogous to the monopole-antimonopole chains 
encountered in the Georgi-Glashow model \cite{kks}.

But monopole-antimonopole systems also exhibit vortex rings,
where the Higgs field vanishes not (only) on isolated points
on the symmetry axis but (also) on one or more rings,
centered around the symmetry axis \cite{kks}.
Here we show, that such vortex ring solutions arise
in Weinberg-Salam theory as well,
when systems of multisphalerons and -antisphalerons
are considered with $n \ge 3$.
We then discuss the properties of these new solutions,
in particular their energies and magnetic dipole moments
\cite{km,multi,klink,hind},
and we consider also the influence of the value of the Higgs mass.

In section II we briefly review the bosonic sector of Weinberg-Salam theory. 
We present the static axially symmetric Ans\"atze 
and the boundary conditions for these new solutions in section III.
In section IV we then present our numerical results
for sphaleron-antisphaleron pairs, chains and vortex rings,
and discuss the physical properties of these solutions.
We give our conclusions in section V.

\section{Weinberg-Salam Lagrangian}

We consider the bosonic sector of Weinberg-Salam theory
\begin{equation}
{\cal L} = -\frac{1}{2} {\rm Tr} (F_{\mu\nu} F^{\mu\nu})
-  \frac{1}{4}f_{\mu \nu} f^{\mu \nu}                                           
- (D_\mu \Phi)^{\dagger} (D^\mu \Phi) 
- \lambda (\Phi^{\dagger} \Phi - \frac{v^2}{2} )^2 
\  
\label{lag1}
\end{equation}
with su(2) field strength tensor
\begin{equation}
F_{\mu\nu}=\partial_\mu V_\nu-\partial_\nu V_\mu
            + i g [V_\mu , V_\nu ]
\ , \end{equation}
su(2) gauge potential $V_\mu = V_\mu^a \tau_a/2$,
u(1) field strength tensor
\begin{equation}
f_{\mu\nu}=\partial_\mu A_\nu-\partial_\nu A_\mu 
\ , \end{equation}
and covariant derivative of the Higgs field
\begin{equation}
D_{\mu} \Phi = \Bigl(\partial_{\mu}
             +i g  V_{\mu} 
             +i \frac{g'}{2} A_{\mu} \Bigr)\Phi
\ , \end{equation}
where $g$ and $g'$ denote the SU(2) and U(1) gauge coupling constants,
respectively,
$\lambda$ denotes the strength of the Higgs self-interaction and
$v$ the vacuum expectation value of the Higgs field.

The Lagrangian (\ref{lag1}) is invariant under local $SU(2)$
gauge transformations $U$,
\begin{eqnarray}
V_\mu &\longrightarrow & U V_\mu U^\dagger
+ \frac{i}{g} \partial_\mu U  U^\dagger \ ,
\nonumber\\
\Phi  &\longrightarrow & U \Phi\ .
\nonumber
\end{eqnarray}
The gauge symmetry is spontaneously broken 
due to the non-vanishing vacuum expectation
value of the Higgs field
\begin{equation}
    \langle \Phi \rangle = \frac{v}{\sqrt2}
    \left( \begin{array}{c} 0\\1  \end{array} \right)   
\ , \end{equation}
leading to the boson masses
\begin{equation}
    M_W = \frac{1}{2} g v \ , \ \ \ \ 
    M_Z = \frac{1}{2} \sqrt{(g^2+g'^2)} v \ , \ \ \ \ 
    M_H = v \sqrt{2 \lambda} \ . 
\end{equation}
$ \tan \theta_w = g'/g $ determines
the weak mixing angle $\theta_w$,
defining the electric charge $e = g \sin \theta_w$.  

In Weinberg-Salam theory, baryon number is not conserved
\begin{equation}
 \frac{d Q_{\rm B}}{dt} = \int d^3 r \partial_t j^0_{\rm B}
= \int d^3 r \left[ \vec \nabla \cdot \vec j_{\rm B}
 + \frac{g^2}{32 \pi^2} \epsilon^{\mu\nu\rho\sigma} \,
{\rm Tr} \left(F_{\mu\nu} F_{\rho\sigma} \right) \right] \ . 
\end{equation}
Starting at time $t=-\infty$ at the vacuum with $Q_{\rm B}=0$,
one obtains the baryon number of a sphaleron solution at
time $t=t_0$ \cite{km},
\begin{equation}
 Q_{\rm B} = 
\int_{-\infty}^{t_0} dt \int_S \vec K \cdot d \vec S
+  \int_{t=t_0} d^3r K^0 \ , 
\end{equation}
where the $\vec \nabla \cdot \vec j_{\rm B}$ term is neglected,
and the anomaly term is reexpressed in terms of the
Chern-Simons current
\begin{equation}
 K^\mu=\frac{g^2}{16\pi^2}\varepsilon^{\mu\nu\rho\sigma} {\rm Tr}(
 F_{\nu\rho}V_\sigma
 + \frac{2}{3} i g V_\nu V_\rho V_\sigma )
\ . \end{equation}
In a gauge, where
\begin{equation}
V_\mu \to \frac{i}{g} \partial_\mu \hat{U} \hat{U}^\dagger \ , \ \ \ 
\hat{U}(\infty) = 1 \ , 
\end{equation}
$\vec K$ vanishes at infinity, yielding for the baryon charge
of a sphaleron solution
\begin{equation}
 Q_{\rm B} = \int_{t=t_0} d^3r K^0 \ .
\label{Q}
\end{equation}

In static classical solutions of the general
field equations the time components of
the gauge fields vanish, $V_0=0$ and $A_0=0$.
For non-vanishing $g'$ it is inconsistent to set the U(1) field to
zero, since the SU(2) gauge field generates a non-vanishing current
\begin{equation}
j_i = -\frac{i}{2} g' (\Phi^\dagger D_i \Phi - (D_i\Phi)^\dagger \Phi) \
\label{u1current}
\end{equation}
which acts as a source for the gauge potential $A_i$.
This current then determines the
magnetic dipole moment $\vec{\mu}$ of a classical configuration, since
\begin{equation}
\vec{\mu} = \frac{1}{2} \int \vec{r} \times \vec{j}\, d^3r \ .
\label{mu}
\end{equation}
When $g'=0$, the U(1) gauge potential $A_\mu$ decouples
and may consistently be set to zero.

Setting the weak mixing angle to zero
is a good approximation for sphalerons and multisphalerons \cite{multi}.
We therefore here construct sphaleron, sphaleron-antisphaleron chain
and vortex ring solutions 
in the limit of vanishing weak mixing angle. 
We determine the magnetic dipole moments of these solutions
perturbatively \cite{km},
since the ratio $\vec{\mu}/e$ remains finite
for $\theta_w \rightarrow 0$.

\section{Ansatz and boundary conditions}

To obtain new classical solutions of Weinberg-Salam theory
(at vanishing weak mixing angle),
we employ the static axially symmetric ansatz
\cite{bk}
\begin{equation}
V_i dx^i = \left(\frac{H_1}{r} dr + (1-H_2) d\theta\right)
           \frac{\tau^{(n)}_\vphi}{2g}
          -n\sin\theta\left(H_3 \frac{\tau^{(n,m)}_r}{2g} 
	  + (1-H_4)\frac{\tau^{(n,m)}_\theta}{2g}\right) d\vphi
	   \ , \ \ \ V_0=0 \ ,
\label{a_axsym}
\end{equation}	  
and
\begin{equation}
\Phi = i( \Phi_1 \tau^{(n,m)}_r + \Phi_2 \tau^{(n,m)}_\theta )\frac{v}{\sqrt2}
    \left( \begin{array}{c} 0\\1  \end{array} \right) \ .
\end{equation}
where
\begin{eqnarray}	  
\tau^{(n,m)}_r & = & \sin m\theta (\cos n\vphi \tau_x + \sin n\vphi \tau_y) 
           + \cos m\theta \tau_z \ , \ \ 
\nonumber \\	   
\tau^{(n,m)}_\theta & = & \cos m\theta (\cos n\vphi \tau_x + \sin n\vphi \tau_y) 
           - \sin m\theta \tau_z \ , \ \ 
\nonumber \\	   
\tau^{(n)}_\vphi & = & (-\sin n\vphi \tau_x + \cos n\vphi \tau_y) 
\ , \ \ \nonumber 
\end{eqnarray}	  
$n$ and $m$ are integers, 
and $\tau_x$, $\tau_y$ and $\tau_z$ denote the Pauli matrices.

The two integers $n$ and $m$ determine the baryon number of the
solutions \cite{bk,Kari},
\begin{equation}
Q_{\rm B}= \frac{n \ (1-(-1)^m)}{4} \, .
\end{equation}
For $m=n=1$ the Ansatz yields the 
Klinkhamer-Manton sphaleron \cite{km}.
For $n>1$ or $m>1$,
the functions $H_1,\dots,H_4$, $\Phi_1$, and $\Phi_2$ 
depend on $r$ and $\theta$, only. 
These axially symmetric solutions represent
multisphaleron configurations \cite{multi}, when $n>1$ and $m=1$.
When $n=1$ and $m>1$, they represent sphaleron-antisphaleron
pairs ($m=2$) \cite{klink}, or sphaleron-antisphaleron chains,
as shown below.

With this Ansatz the full set of field equations reduces to a system 
of six coupled partial differential equations in the independent variables 
$r$ and $\theta$. A residual U(1) gauge degree of freedom is 
fixed by the condition $r\partial_r H_1 - \partial_\theta H_2=0$ \cite{kkb}.

Regularity and finite energy require the boundary conditons 
\begin{eqnarray}
%{\rm at }\ 
r=0: &  & \hspace{1.0cm}
H_1=H_3=0\, , \ H_2=H_4=1\, , \ \hspace{4.13cm}
\Phi_1=\Phi_2=0 \, , 
\nonumber \\
r\rightarrow \infty: &  &  \hspace{1.0cm}
H_1=H_3=0\, , \ H_2=1-2m \, ,  \ 
                       1-H_4=\frac{2\sin m\theta}{\sin\theta} 
\, , \ \hspace{1.0cm}
\Phi_1= 1 \, , \ \Phi_2=0 \, ,  
\nonumber \\
\theta = 0,\pi: &  & \hspace{1.0cm} 
H_1=H_3=0\, ,  \ \partial_\theta H_2=\partial_\theta H_4=0 
\, , \ \hspace{3.43cm}
\partial_\theta \Phi_1=0  \, , \Phi_2=0 \, ,  
\end{eqnarray}
for odd $m$, while at $r=0$
$ (\sin m\theta \Phi_1+\cos m\theta \Phi_2) = 0 $, 
$ \partial_r (\cos m\theta \Phi_1-\sin m\theta \Phi_2) = 0 $ 
is required for even $m$.

\section{Results}

We first briefly recall the static axially symmetric 
multisphaleron solutions, emerging for $m=1$ and $n>1$ \cite{multi}.
These solutions represent superpositions of $n$ sphalerons,
where the modulus of the Higgs field has a single
isolated node, located at the origin.
The energy density is not maximal at the origin, however,
but concentrated in a torus-like region centered around
the symmetry axis at the origin. %and w.r.t.~the $xy$-plane. 
With increasing $n$ the maximum of the energy density
moves outward, i.e., the associated torus increases in radius.
The magnetic dipole moment of the multisphalerons
rises almost linearly with $n$, as expected for such a superposition
\cite{multi}.

Let us now turn to 
the sphaleron-antisphaleron pair with $m=2$ and $n=1$ \cite{klink}.
The modulus of the Higgs field of this configuration
has two isolated nodes on the symmetry axis,
where the centers of the sphaleron and the antisphaleron are located,
respectively.
The distance $d_{(m,n)}$ between the nodes shrinks slightly, 
when the Higgs mass
is increased from zero, reaching a miminum of 
$d_{(2,1)} \approx 4 M_{\rm W}^{-1}$ at 
$M_{\rm H}/M_{\rm W} \approx 0.3$. The distance $d_{(2,1)}$
then increases to $d_{(2,1)} \approx 9.2 M_{\rm W}^{-1}$ 
at $M_{\rm H}/M_{\rm W}=1$ \cite{footnote}.

A surface of large constant energy density 
of such a sphaleron-antisphaleron pair
then consists of two sphere-like surfaces, 
surrounding these two nodes.
The energy density and the modulus of the Higgs field 
are both exhibited in Fig.~\ref{f-1}
for $M_{\rm H}/M_{\rm W}=1$.

\begin{figure}[h!]
\lbfig{f-1}
\begin{center}
\hspace{0.0cm} (a)\hspace{-0.6cm}
\includegraphics[height=.25\textheight, angle =0]{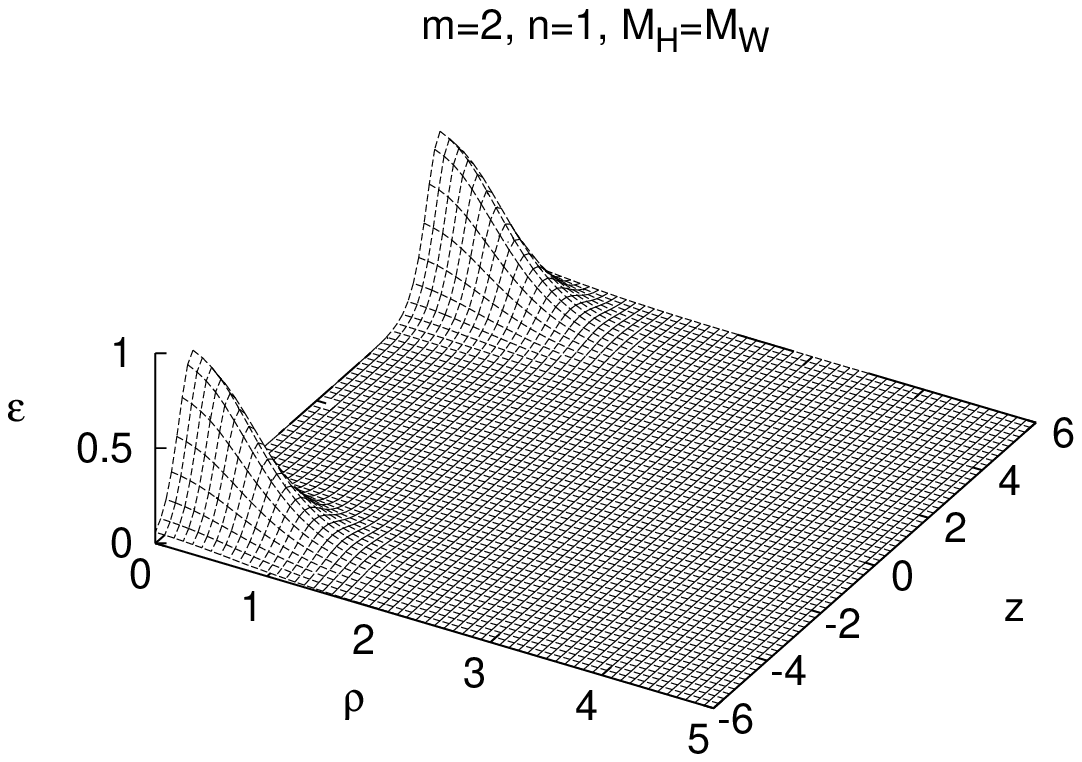}
\hspace{0.5cm} (b)\hspace{-0.6cm}
\includegraphics[height=.25\textheight, angle =0]{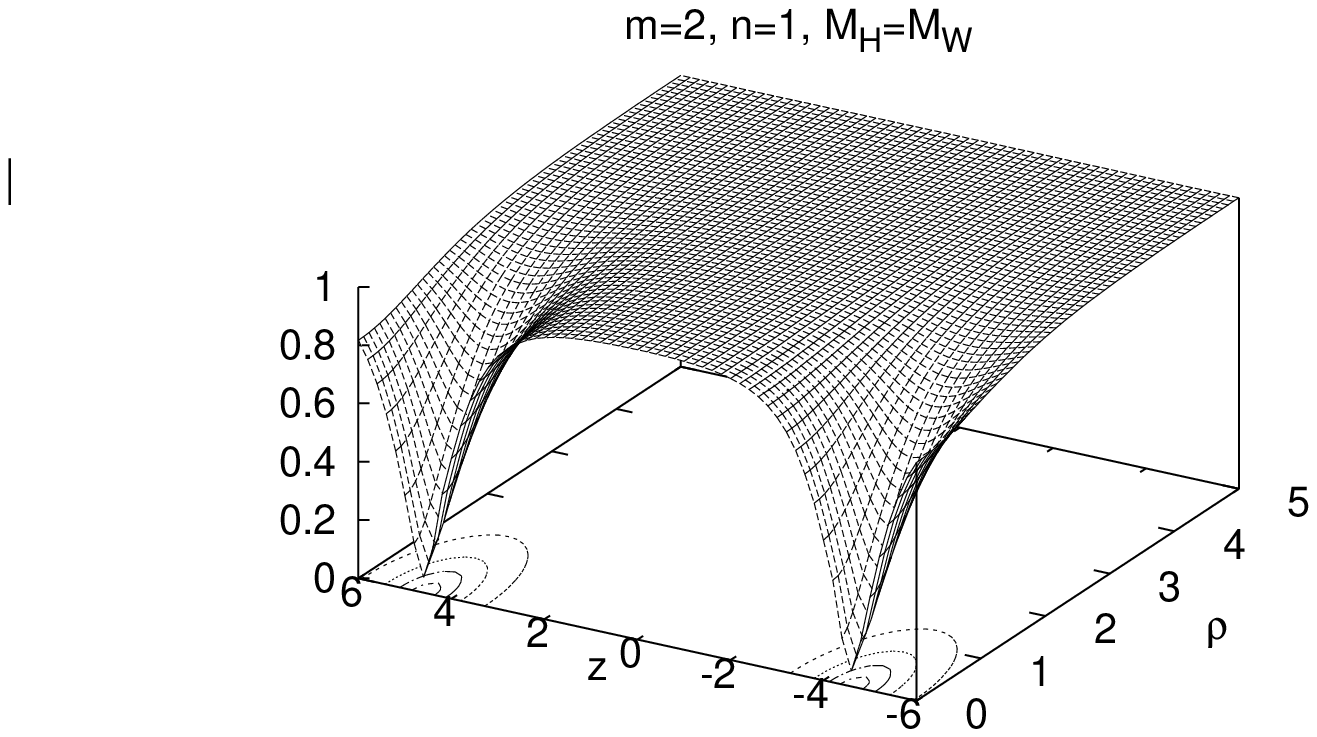}
\\
\hspace{0.0cm} (c)\hspace{-0.6cm}
\includegraphics[height=.25\textheight, angle =0]{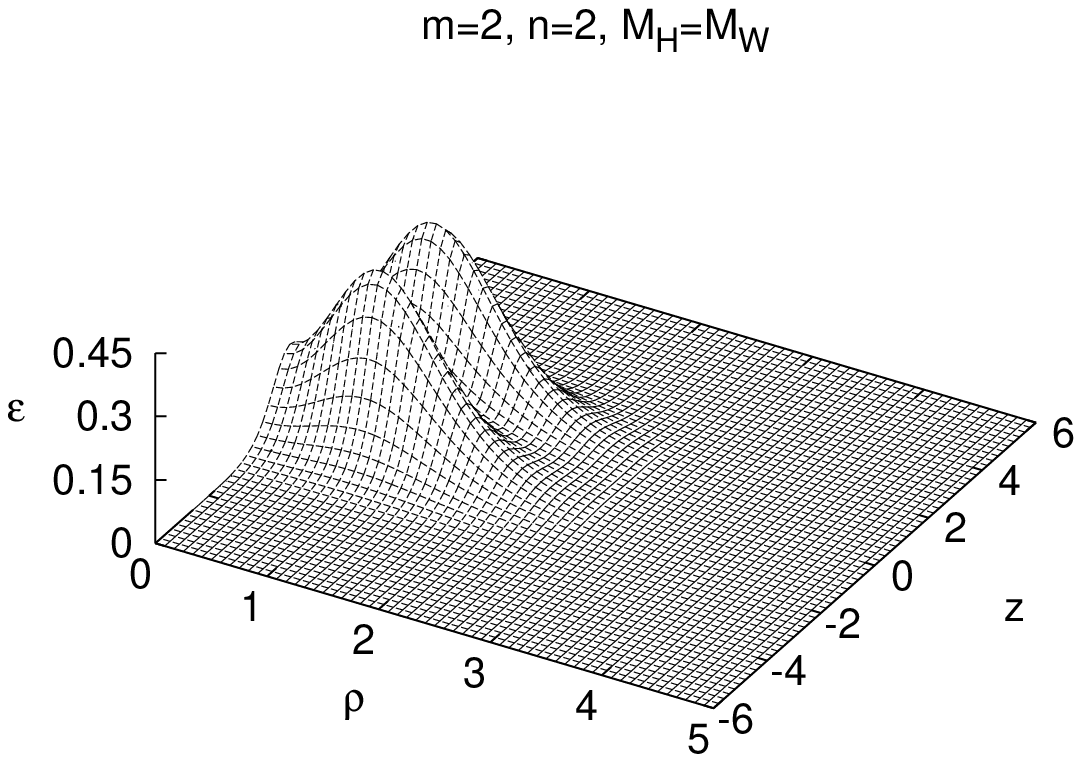}
\hspace{0.5cm} (d)\hspace{-0.6cm}
\includegraphics[height=.25\textheight, angle =0]{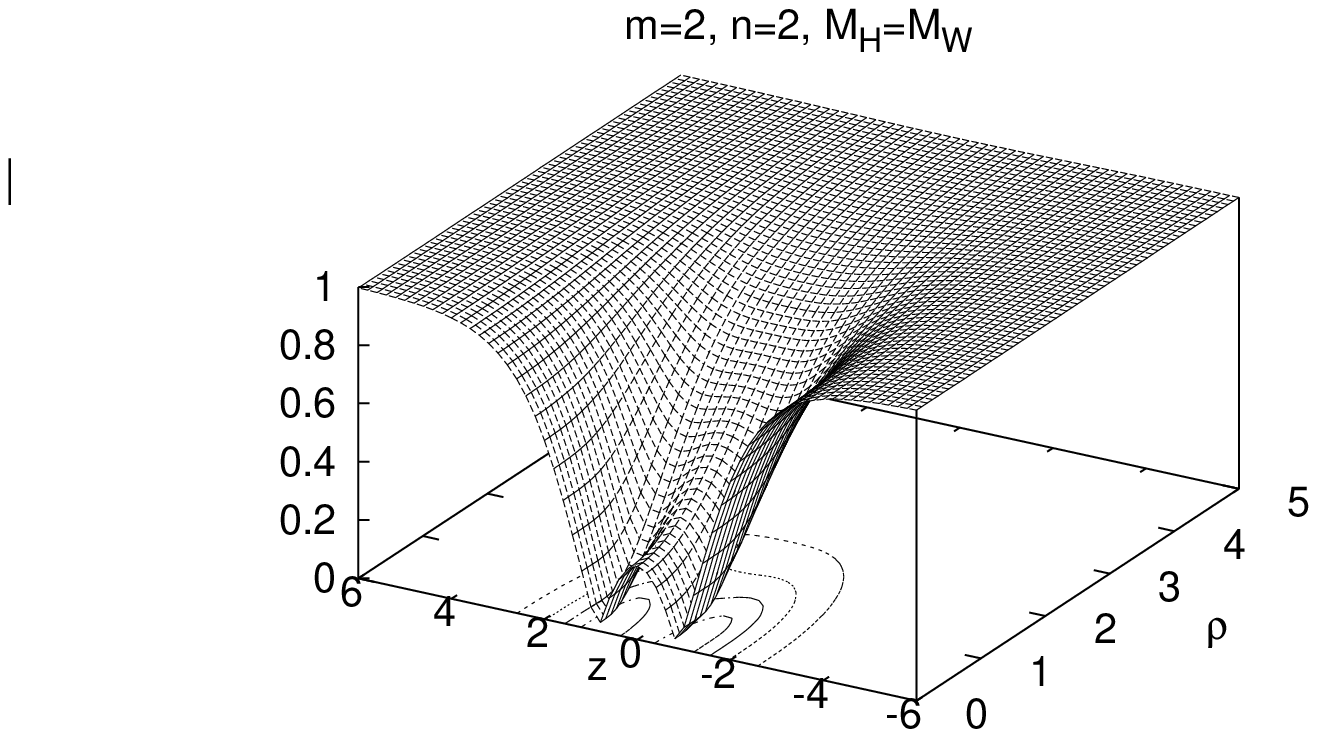}
\\
\hspace{0.0cm} (e)\hspace{-0.6cm}
\includegraphics[height=.25\textheight, angle =0]{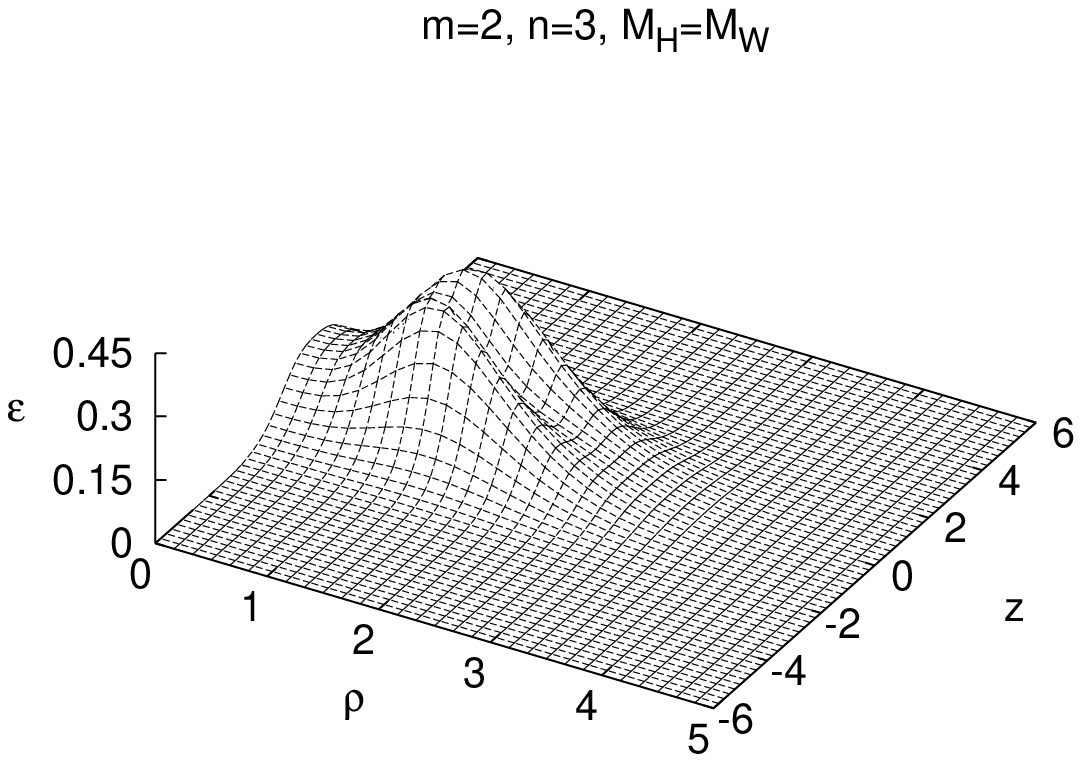}
\hspace{0.5cm} (f)\hspace{-0.6cm}
\includegraphics[height=.25\textheight, angle =0]{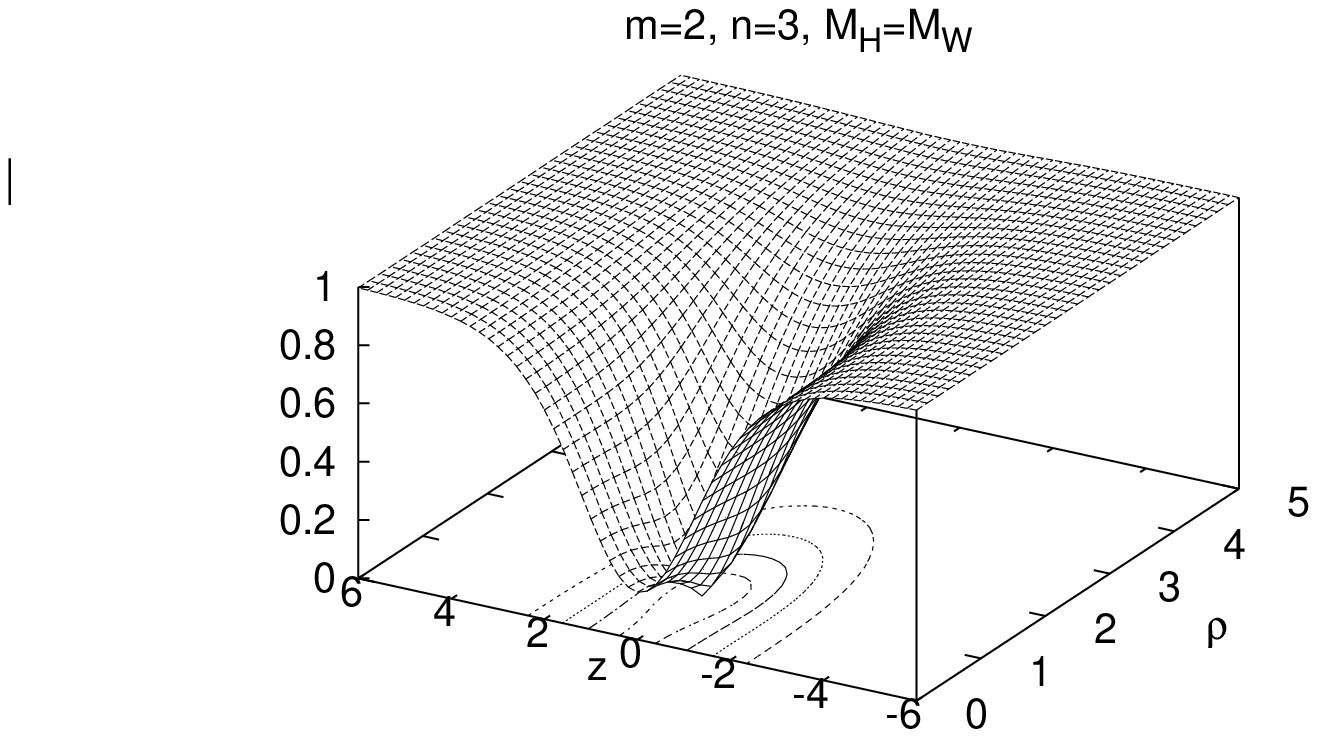}
\\
\hspace{0.0cm} (g)\hspace{-0.6cm}
\includegraphics[height=.25\textheight, angle =0]{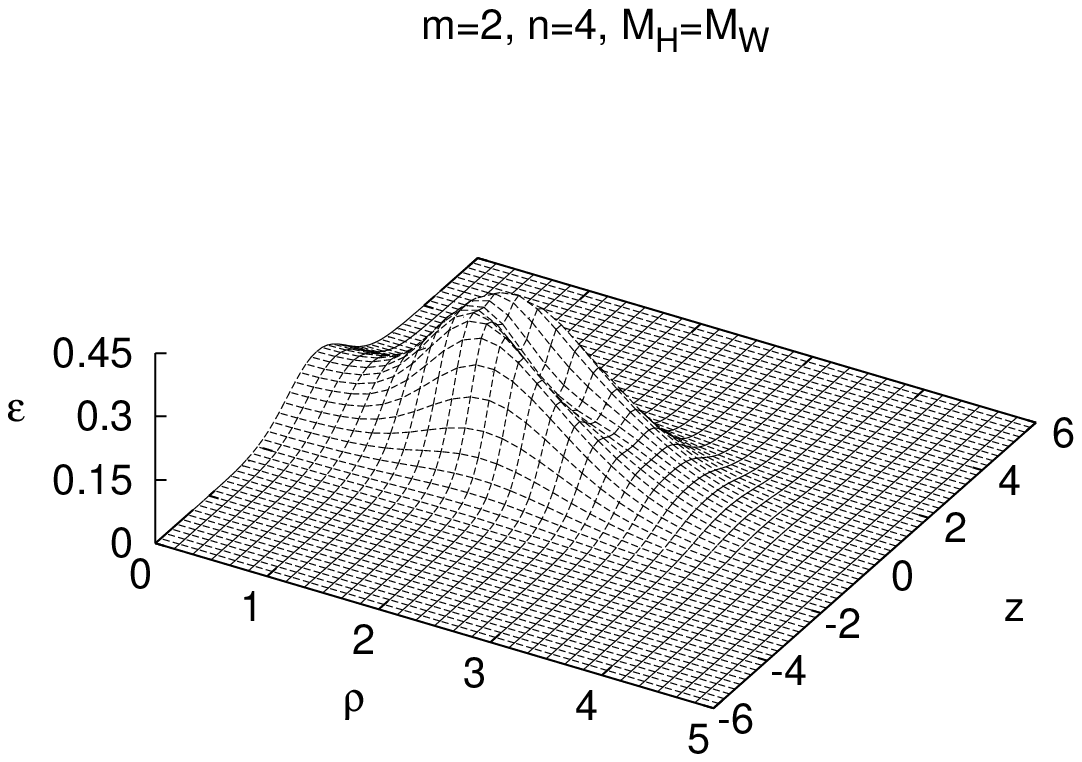}
\hspace{0.5cm} (h)\hspace{-0.6cm}
\includegraphics[height=.25\textheight, angle =0]{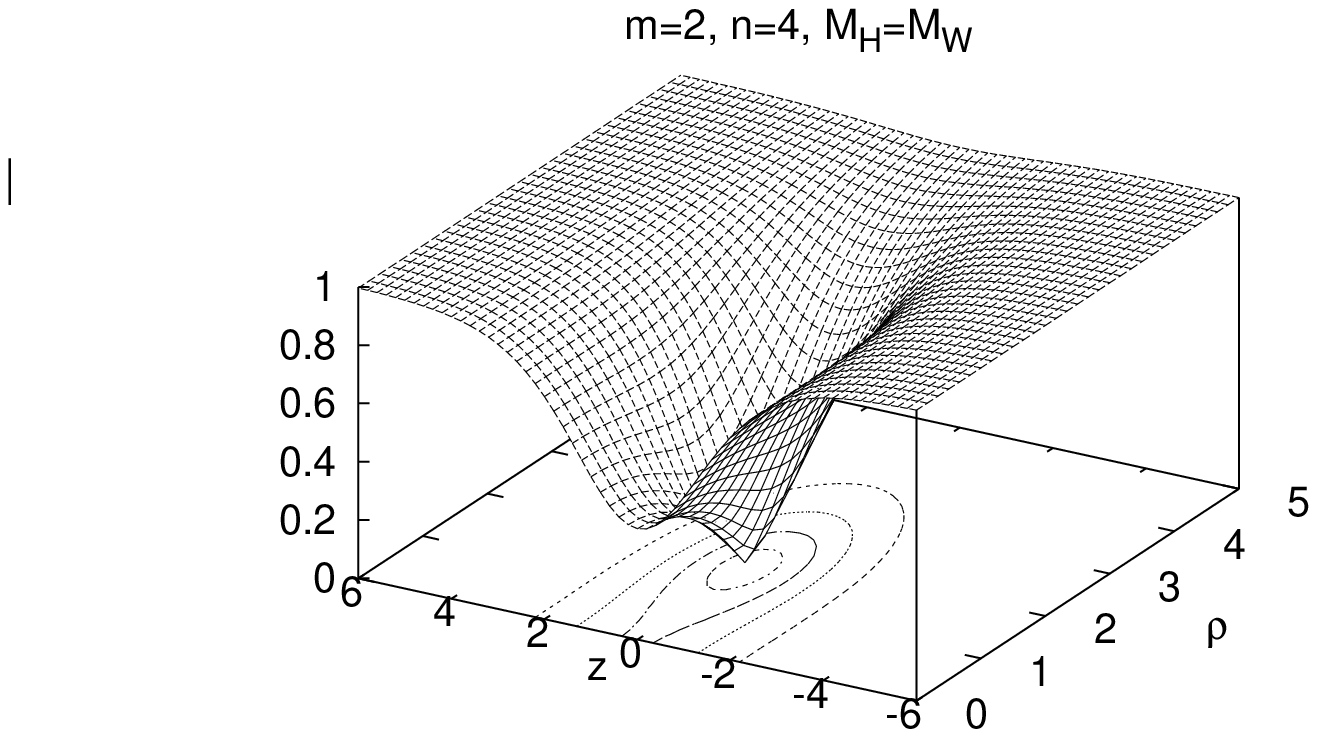}
\end{center}
\vspace{-0.5cm}
\caption{\small
The energy density $\varepsilon$ (left) and the modulus of the Higgs field
$|\Phi|$ (right) are exhibited 
(versus the coordinates $\rho$ and $z$ in units of $M_{\rm W}^{-1}$)
for sphaleron-antisphaleron solutions ($m=2$, $n=1$)
and their multisphaleron generalizations ($m=2$, $n=2-4$) 
for $M_{\rm H}/M_{\rm W}=1$.
}
\end{figure}

The mass $E_{(m,n)}$ of the sphaleron-antisphaleron solutions 
depends on the value of the Higgs mass
and thus the strength of the scalar self-interaction.
It is then of particular interest to see, whether the
pair is energetically bound.
For small Higgs masses this is clearly the case,
since the binding energy is quite large.
In the case of vanishing Higgs mass, for instance,
$E_{(2,1)}/E_{(1,1)}=1.877$.
But for $M_{\rm H}/M_{\rm W}=1$ the binding energy
has become quite small,
$E_{(2,1)}/E_{(1,1)}\approx 1.997\pm 0.003$ \cite{footnote}.
For large values of the Higgs mass
the pair may no longer form an energetically bound system,
but we were not able to confirm this due to the numerical inaccuracies
encountered \cite{footnote}.

The sphaleron-antisphaleron pair possesses a magnetic dipole moment
$\mu_{(m,n)}$.
Its magnitude is about twice the magnitude of the magnetic dipole moment of a 
single sphaleron, 
since the magnetic dipole moments of the sphaleron and the antisphaleron 
in the pair add up.
For instance, for $M_{\rm H}/M_{\rm W}=1$ their ratio is
$\mu_{(2,1)}/\mu_{(1,1)}\approx 2.0$.
Considering the dependence of the magnitude of the magnetic dipole moment 
on the Higgs mass, we observe a decrease by roughly one quarter 
as the Higgs mass increases from zero to $M_{\rm H}/M_{\rm W}=1$. 

Let us now consider how the configurations change as $n$ is increased.
For $n>1$, these sphaleron-antisphaleron pair solutions 
are expected to be composed of a multisphaleron and a 
corresponding antisphaleron.
For $n=2$ this is indeed the case.
But the nodes of the $n=2$ sphaleron and antisphaleron
are located much closer to each other on the symmetry axis
as compared to the $n=1$ pair.
For instance, for $M_{\rm H}/M_{\rm W} =1$ the distance of the nodes
has decreased to $d_{(2,2)} = 1.7 M_{\rm W}^{-1}$.

The energy density of the $n=2$ pair
no longer has its maxima on the symmetry axis. 
Instead the maxima form rings, located symmetrically
above and below the $xy$-plane, as seen in Fig.~\ref{f-1}.
The energy density (for large values) therefore exhibits
a double torus-like shape.
This conforms to the expectation,
since for a single $n=2$ sphaleron the energy density is torus-like.

The $n=2$ sphaleron-antisphaleron pair is energetically bound.
In the case of vanishing Higgs mass, for instance,
$E_{(2,2)}/E_{(1,2)}=1.756$
resp.~$E_{(2,2)}/E_{(1,1)}=3.175$,
where for $M_{\rm H}/M_{\rm W}=1$ the binding energy
is again much smaller,
$E_{(2,2)}/E_{(1,2)}=1.960$
resp.~$E_{(2,2)}/E_{(1,1)}=3.877$.

The magnetic dipole moment of the $n=2$ sphaleron-antisphaleron pair 
has a magnitude of about twice the magnitude of the magnetic dipole moment 
of an $n=2$ (multi)sphaleron resp.~four times the magnitude of
an $n=1$ sphaleron.
For instance, for $M_{\rm H}/M_{\rm W}=1$, the ratio of
magnetic dipole moments is
$\mu_{(2,2)}/\mu_{(1,1)}=3.721$.

Increasing $n$ further, however, leads to solutions of a completely different
character. Here the modulus of the Higgs field has no longer
two isolated nodes, located on the symmetry axis.
Instead the modulus of the Higgs field vanishes on a ring,
located in the $xy$-plane, when $n \ge 3$.
(Our solutions comprise $n=1,\dots,8$).
Therefore, we refer to these solutions no longer as
sphaleron-antisphaleron pair solutions, but 
sphaleron-antisphaleron vortex rings.
With increasing $n$, the radius $r_{(m,n)}$ of the nodal ring then increases
(almost linearly for $n\ge 4$),
from $r_{(2,3)}= 1.0 M_{\rm W}^{-1}$ to $r_{(2,8)}=3.4 M_{\rm W}^{-1}$
for $M_{\rm H}/M_{\rm W}=1$.

In contrast to the single nodal ring of the
modulus of the Higgs field, present for $n \ge 3$,
the energy density still exhibits
two tori, located symmetrically above and below the $xy$-plane,
when $n=3$ and $4$, as seen in Fig.~\ref{f-1}.
But for $n \ge 5$ the structure of the energy density conforms to the 
nodal structure, exhibiting a single central torus,
which increases in size with increasing $n$.

The mass of the solutions increases approximately linearly with $n$,
as well. This is seen in Fig.~\ref{f-2}, where the scaled mass
$E_{(m,n)}/(m E_{(1,1)})$ of the solutions
is shown versus $n$ for $M_{\rm H}/M_{\rm W}=1$.
Whereas the multisphalerons (where $m=1$) are already unbound \cite{multi}
at this value of the Higgs mass, the 
$m=2$ sphaleron-antisphaleron vortex rings are clearly energetically bound.

The magnetic dipole moment $\mu_{(2,n)}$ of the solutions 
rises approximately linearly with $n$ % x**1.2
for vanishing Higgs mass, and slightly faster
for finite Higgs mass. % x**1.3
The magnetic dipole moment is also exhibited in Fig.~\ref{f-2}
for $M_{\rm H}/M_{\rm W}=1$.

\begin{figure}[h!]
\lbfig{f-2}
\begin{center}
\hspace{0.0cm} \hspace{-0.6cm}
\includegraphics[height=.25\textheight, angle =0]{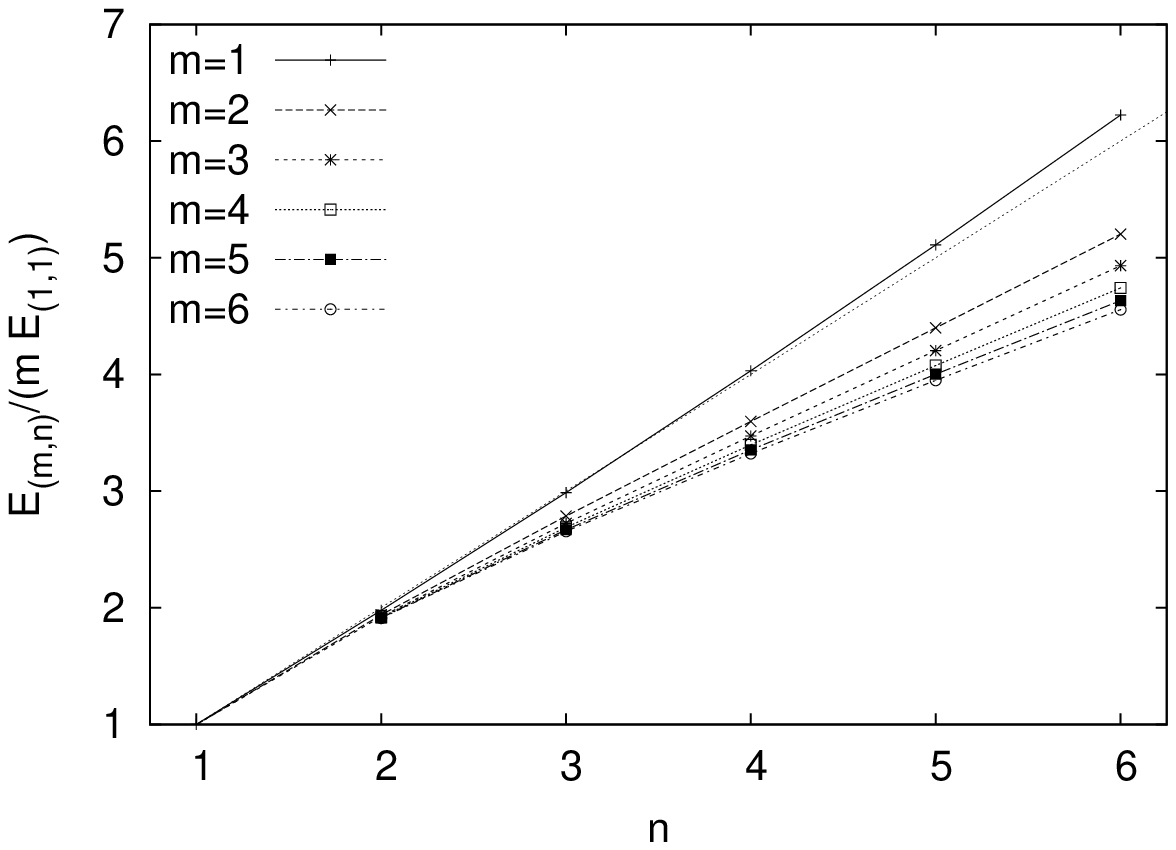}
\hspace{0.5cm} \hspace{-0.6cm}
\includegraphics[height=.25\textheight, angle =0]{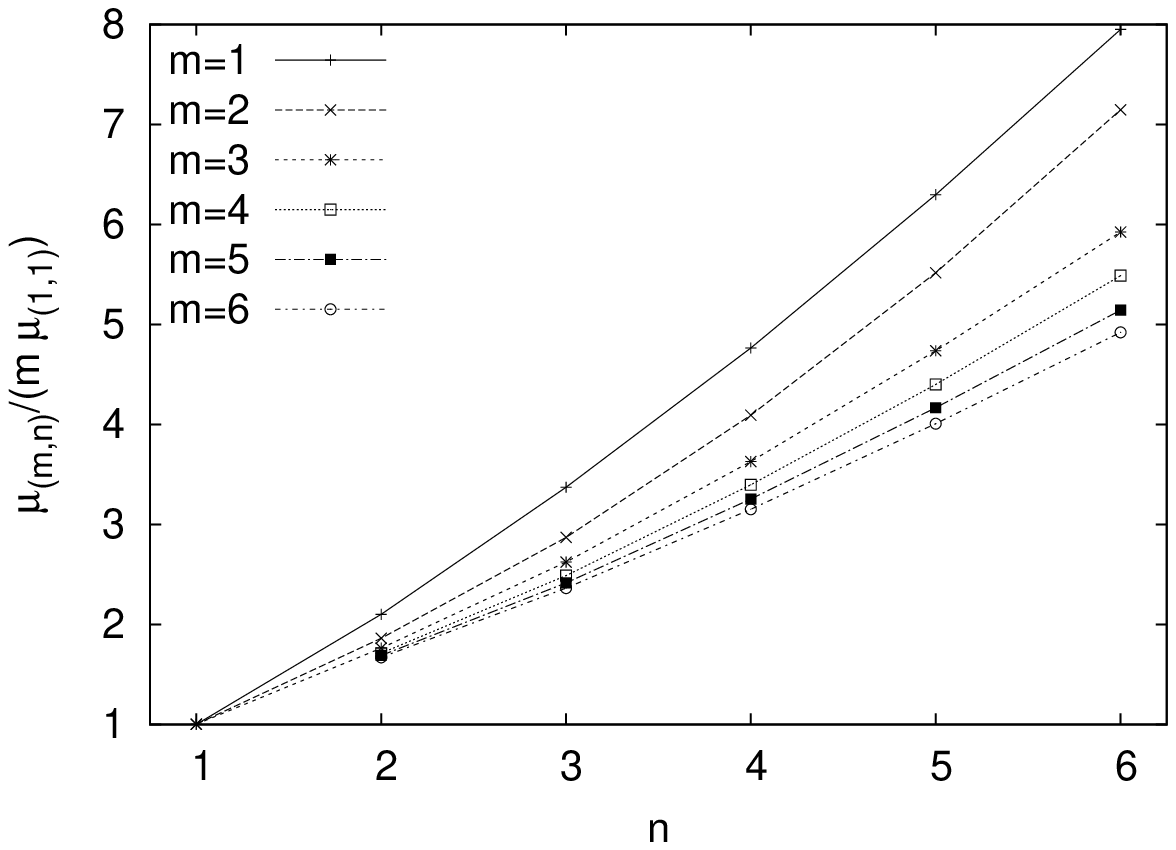}
\end{center}
\vspace{-0.5cm}
\caption{\small
Scaled mass $E_{(m,n)}/(m E_{(1,1)})$ (left)
and scaled magnetic dipole moment $\mu_{(m,n)}/(m \mu_{(1,1)})$ (right)
of sphaleron-antisphaleron systems with $m=1,\dots,6$, $n=1,\dots,6$
for $M_{\rm H}/M_{\rm W}=1$.
The solutions below the thin dotted line (left) are energetically bound.
}
\end{figure}

We next consider the structure of the solutions
obtained when the integer $m$ is increased.
For $m=3$ and $n=1$ we expect a sphaleron-antisphaleron-sphaleron
configuration.
Indeed, the modulus of the Higgs field of this configuration
exhibits $3$ isolated nodes on the symmetry axis, with large separation, 
e.g., $d_{(3,1)}\approx 10 M_{\rm W}^{-1}$ for $M_{\rm H}/M_{\rm W}=1$
\cite{footnote}.
At the same time the energy density is maximal in the vicinity of these nodes.
Thus surfaces of large energy density form a small chain of
$3$ sphere-like surfaces, located along the symmetry axis.

For $m=3$ and $n=2$, the modulus of the Higgs field
still possesses three isolated nodes on the symmetry axis,
but their mutual distance has strongly decreased, to about $1.9 M_{\rm W}^{-1}$. 
As expected, the energy density now forms $3$ tori,
as seen in Fig.~\ref{f-3} for $M_{\rm H}/M_{\rm W}=1$.

\begin{figure}[h!]
\lbfig{f-3}
\begin{center}
\hspace{0.0cm} (a)\hspace{-0.6cm}
\includegraphics[height=.25\textheight, angle =0]{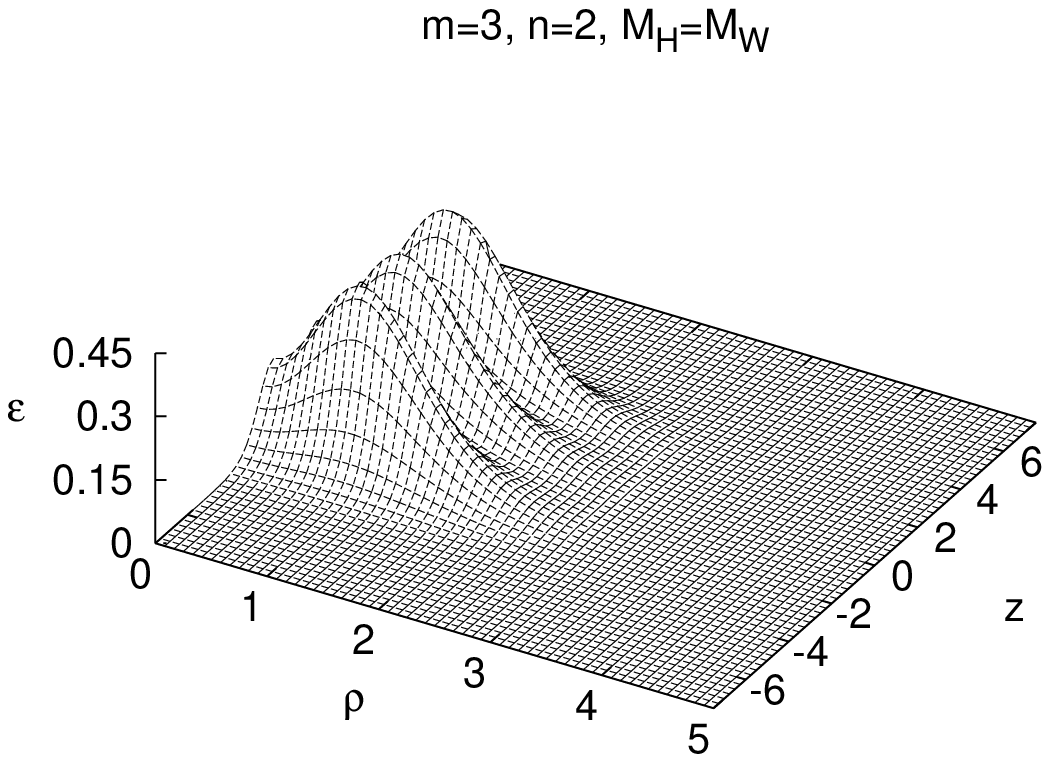}
\hspace{0.5cm} (b)\hspace{-0.6cm}
\includegraphics[height=.25\textheight, angle =0]{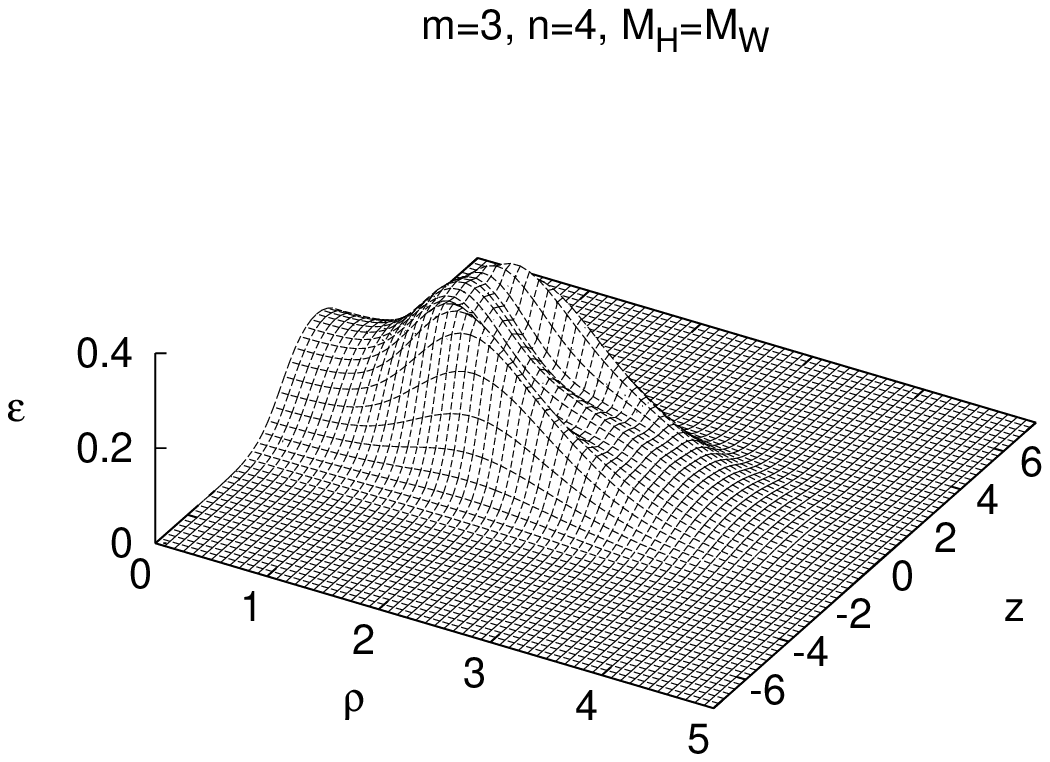}
\\
\hspace{0.0cm} (c)\hspace{-0.6cm}
\includegraphics[height=.25\textheight, angle =0]{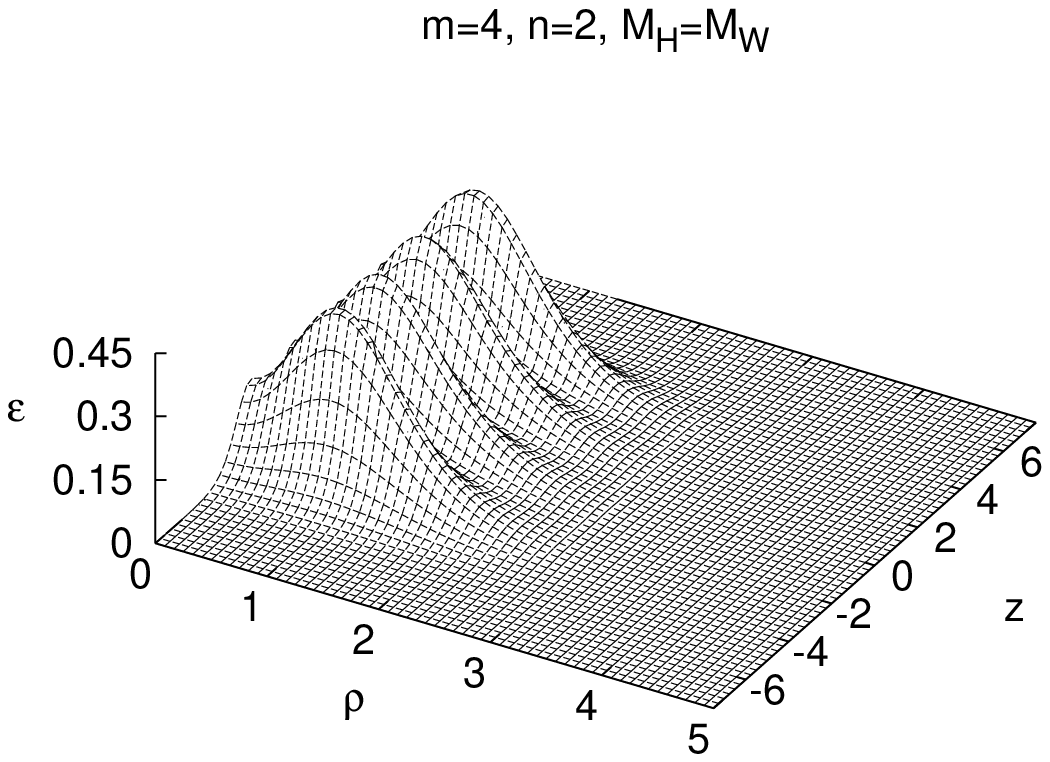}
\hspace{0.5cm} (d)\hspace{-0.6cm}
\includegraphics[height=.25\textheight, angle =0]{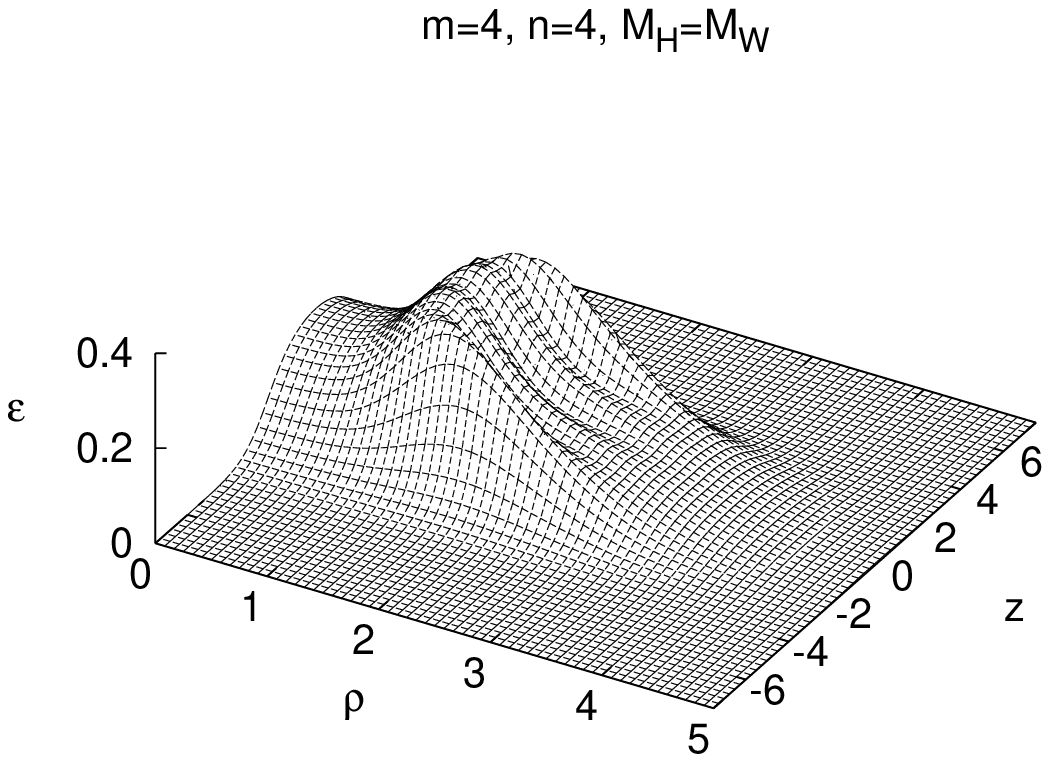}
\\
\hspace{0.0cm} (e)\hspace{-0.6cm}
\includegraphics[height=.25\textheight, angle =0]{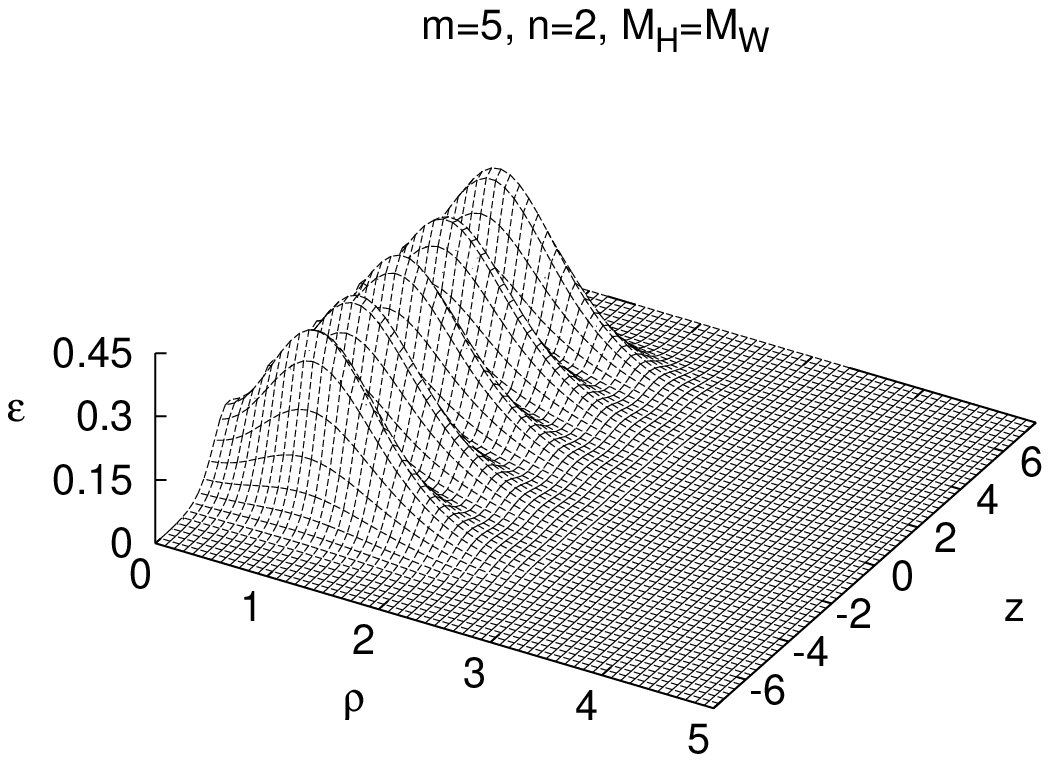}
\hspace{0.5cm} (f)\hspace{-0.6cm}
\includegraphics[height=.25\textheight, angle =0]{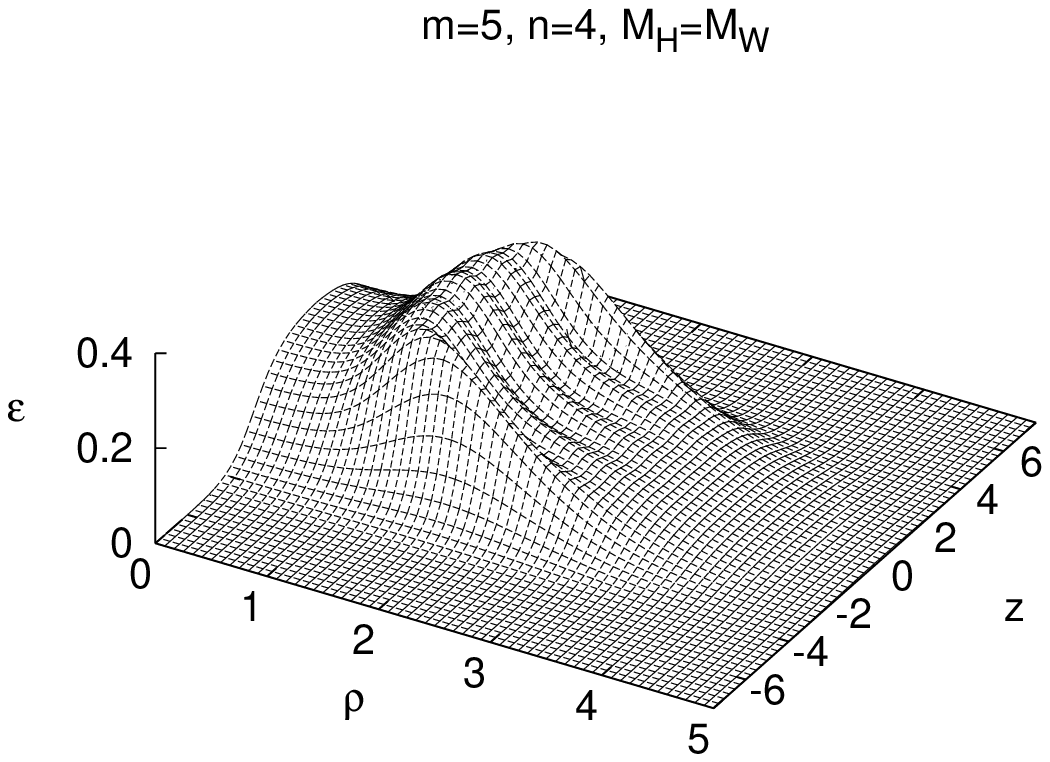}
\\
\hspace{0.0cm} (g)\hspace{-0.6cm}
\includegraphics[height=.25\textheight, angle =0]{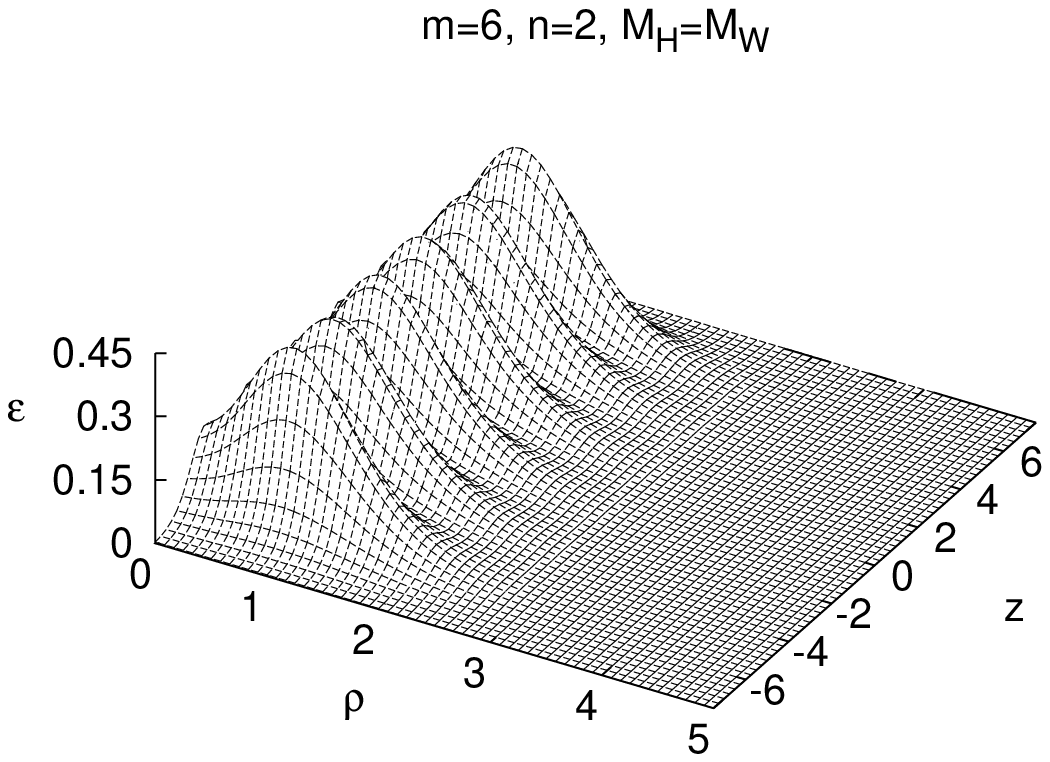}
\hspace{0.5cm} (h)\hspace{-0.6cm}
\includegraphics[height=.25\textheight, angle =0]{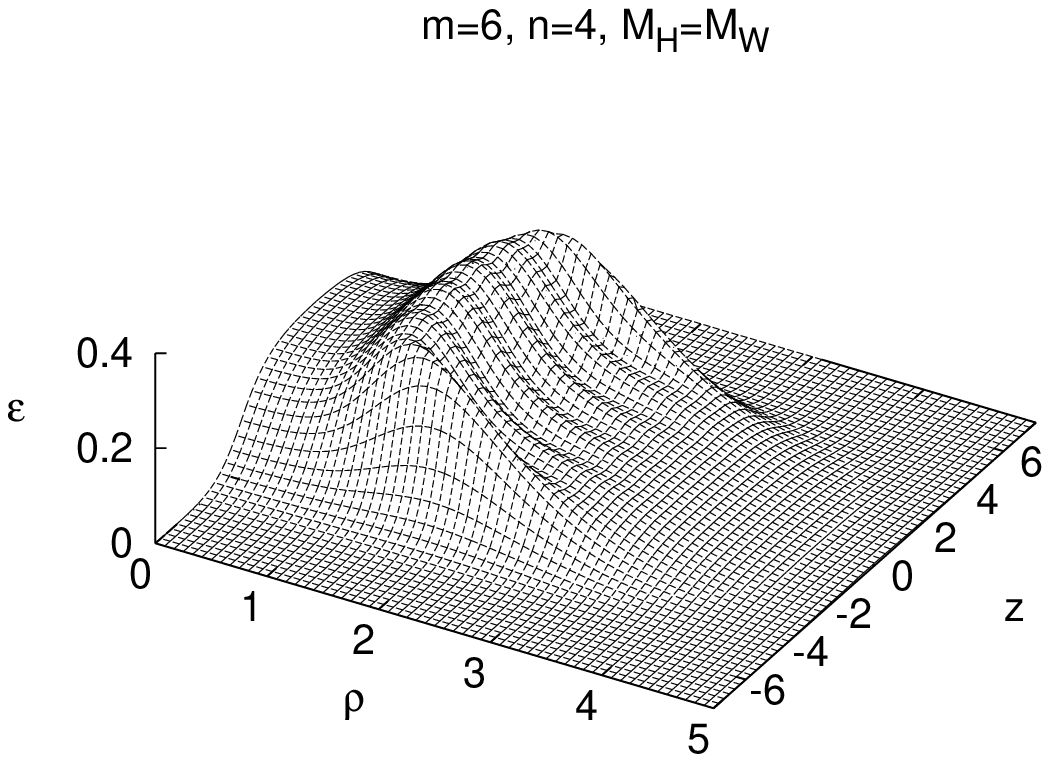}
\end{center}
\vspace{-0.5cm}
\caption{\small
The energy density $\varepsilon$ 
% and the modulus of the Higgs field $|\Phi|$
is exhibited 
(versus the coordinates $\rho$ and $z$ in units of $M_{\rm W}^{-1}$)
for sphaleron-antisphaleron chain (left: $m=3-6$, $n=2$)
and vortex ring (right: $m=3-6$, $n=4$) solutions
for $M_{\rm H}/M_{\rm W}=1$.
}
\end{figure}

As $n$ increases further, again the character of the solutions
changes, and new configurations appear,
where the modulus of the Higgs field vanishes on rings
centered around the symmetry axis.
The precise evolution of the nodes and
vortex rings with $n$ is, however, sensitive
to the value of the Higgs mass.
For $M_{\rm H}/M_{\rm W}=1$,
and $n=3$ we still observe $3$
isolated nodes on the symmetry axis,
but in addition two rings appear,
which are located symmetrically above and below the $xy$-plane.
For $n=4$ only a single isolated node is left at the origin.
But in addition to the two rings above and below
the $xy$-plane, a third ring has appeared in the $xy$-plane.
For still larger values of $n$, these three vortex rings
approach each other and merge, such that a single
vortex ring in the $xy$-plane is left.
With increasing $n$, the radius $r_{(m,n)}$ of the central nodal ring 
again increases (almost linearly for $n\ge 4$).
For $M_{\rm H}/M_{\rm W}=1$,
the energy density of the $n=4$ vortex ring solution
is also exhibited in Fig.~\ref{f-3}.

It is now conceivable, how the pattern of configurations is going to
continue for the larger values of $m$.
For $n=1$ and $2$, sphaleron-antisphaleron chains arise,
where a total of $m$ sphalerons and antisphalerons are located
alternatingly on the symmetry axis, their locations being revealed
by the $m$ nodes of the modulus of the Higgs field.
For $n=1$ the nodes are farther apart from each other,
while they get closer to each other for $n=2$.
The energy density of the $n=2$ chains with $m=4-6$ is also exhibited in
Fig.~\ref{f-3} for $M_{\rm H}/M_{\rm W}=1$.

For the larger values of $n$, vortex ring solutions arise again.
For $m=4$ the modulus of the Higgs field forms two vortex rings,
located symmetrically above and below the $xy$-plane.
For $m=5$, the modulus of the Higgs field forms likewise two vortex rings,
which, however, are supplemented by an isolated node at the origin.
For $m=6$ the modulus of the Higgs field 
forms two vortex rings for $n=3$, retaining the inner $2$
isolated nodes on the symmetry axis, while for $n>3$ 
only three vortex rings are present,
located symmetrically above, below and in the $xy$-plane, 
%this also holds for $m=7$, 
and so on.

Thus we conjecture, that for the larger values of $n$,
the solutions will possess $[m/2]$ vortex rings
(where $[m/2]$ denotes the integer part of $m/2$).
In the transitional region $n=3$ or $4$,
the configurations may still retain isolated nodes on
the symmetry axis away from the origin
and/or possess a smaller or larger number of vortex rings.
The size of the vortex rings always increases with $n$.
The energy density of the $n=4$ vortex ring solutions with $m=4-6$ 
is also exhibited in Fig.~\ref{f-3} for $M_{\rm H}/M_{\rm W}=1$.

The mass $E_{(m,n)}$ of these sets of solutions and their
magnetic dipole moments $\mu_{(m,n)}$ are exhibited in Fig.~\ref{f-2}
for $M_{\rm H}/M_{\rm W}=1$.
As seen in the figure,
these sphaleron-antisphaleron systems represent clearly
energetically bound solutions for the larger values of $m$ and $n$,
when $M_{\rm H}/M_{\rm W}=1$,
since $E_{(m,n)} < m n E_{(1,1)}$.
Their magnetic dipole moments are to a large extent additive,
although the simple estimate 
$\mu_{(m,n)}=m n \mu_{(1,1)}$ is oversimplified,
at least for the larger values of the Higgs mass, 
such as $M_{\rm H}/M_{\rm W}=1$.

The sphaleron-antisphaleron chain and vortex ring configurations
presented above can all be obtained by first solving for vanishing Higgs mass,
and then successively increasing it. As a function of the Higgs mass 
these solutions then form branches, the fundamental branches.
As observed before for monopole-antimonopole systems \cite{kks},
also here at critical values of the Higgs mass bifurcations arise, 
where new branches of solutions appear. 
For a given set of parameters, the solutions are then no longer unique.
These bifurcations and the additional branches of solutions 
will be discussed elsewhere.

\section{Conclusions}

Concluding, we have found new static axially symmetric solutions of
Weinberg-Salam theory (in the limit of vanishing weak mixing angle),
characterized by two integers, $m$ and $n$,
and carrying baryon number $Q_{\rm B}=n(1-(-1)^m)/4$.
The sphaleron corresponds to the special case $m=n=1$, $Q_{\rm B}=1/2$,
while the sphaleron-antisphaleron pair corresponds to $m=2,n=1$, $Q_{\rm B}=0$.
The new solutions may be considered to fall into two classes,
namely those that mediate baryon number violating transitions and those
that do not \cite{axenides}.
The sphaleron and the sphaleron-antisphaleron pair then
form the lowest energy representatives of these two classes.
Like these representatives the new solutions are unstable,
and correspond to saddle points of the energy functional.

%In the case of vanishing Higgs mass, and 
For $n \le 2$, the modulus of the Higgs field 
of these solutions vanishes on $m$ discrete points
on the symmetry axis, thus these solutions correspond
to sphaleron-antisphaleron pairs and chains.
For $n>2$ the solutions change character,
and the modulus of the Higgs field
vanishes on one or more rings centered around the symmetry axis.
We conjecture, that for large $n$,
the solutions possess $[m/2]$ vortex rings
(where $[m/2]$ denotes the integer part of $m/2$).
For even $m$, these solutions are thus electroweak vortex ring solutions.
Since for odd $m$, one isolated node is retained at the origin,
these solutions represent electroweak sphaleron-vortex ring superpositions.

The mass $E_{(m,n)}$ of these new solutions increases with 
with increasing Higgs mass.
For small and intermediate Higgs mass, the solutions are energetically
bound, $E_{(m,n)} < n m E_{(1,1)}$, but their binding energy
decreases with increasing Higgs mass.
Like the sphaleron, the solutions also possess magnetic dipole moments,
$\mu_{(m,n)}$.
These magnetic dipole moments are to a large extent additive,
although the simple estimate
$\mu_{(m,n)}=m n \mu_{(1,1)}$ holds only approximately.

Based on the small influence of the finite value of the
weak mixing angle on the Klinkhamer-Manton sphaleron and
the multisphalerons \cite{kkb}, we expect, that
the properties of the sphaleron-antisphaleron chain and vortex ring
configurations presented here will be only slightly influenced,
when the physical value of the weak mixing angle will be taken 
into account.

For larger values of the Higgs mass,
the solutions are no longer uniquely specified by the
parameters. Instead bifurcations appear, giving rise to
further configurations, which will be presented elsewhere.
For very large values of the Higgs mass further solutions,
the bisphalerons or `deformed' sphalerons \cite{bi}, are known for $m=n=1$.
These solutions possess less symmetry than the Klinkhamer-Manton sphaleron,
since they do not exhibit parity reflection symmetry.
We expect, that these bispaleron solutions can also be generalized
to axially symmetric configurations, analogous to the ones
presented here. Thus for very large Higgs masses,
bisphaleron chains and vortex ring solutions
may enrich the configuration space even further.
$$ $$

{\bf Acknowledgement}:

B.K.~gratefully acknowledges support by the DFG.


\begin{thebibliography}{000}
\bibitem{footnote}
Note, that solutions with $n=1$, $m>1$ and $M_{\rm H}/M_{\rm W} \gtrapprox 1$
suffer from numerical inaccuracies. 
While uncertainties for the masses and the magnetic moments are only
about $0.2 \%$, numbers for the distances between the nodes
of these  $n=1$, $m>1$ solutions should only be 
regarded as indicative.
\end{thebibliography}
\end{document}